\documentclass[aps,prl,showpacs,twocolumn,preprintnumbers,superscriptaddress,nofootinbib,floatfix,longbibliography,10pt]{revtex4-1}
\usepackage{amsfonts,amssymb,stmaryrd,latexsym,amsmath,braket}
\usepackage{lipsum, babel}
\usepackage{graphicx,subfigure}
\usepackage{comment}
\usepackage{times}
\usepackage{slashed}
\usepackage{bm}
\usepackage{braket}

\makeatletter
\let\saved@includegraphics\includegraphics
\AtBeginDocument{\let\includegraphics\saved@includegraphics}

\makeatother

\begin{document}

\title{Trimmed Sampling Algorithm for the Noisy Generalized Eigenvalue Problem}
\author{Caleb Hicks} 
\author{Dean Lee}
\affiliation{Facility for Rare Isotope Beams and Department of Physics and
Astronomy,
Michigan State University, MI 48824, USA}
\begin{abstract}
Solving the generalized eigenvalue problem is a useful method for finding energy eigenstates of large quantum systems.  It uses projection onto a set of basis states which are typically not orthogonal.  One needs to invert a matrix whose entries are inner products of the basis states, and the process is unfortunately susceptible to even small errors.   The problem is especially bad when matrix elements are evaluated using stochastic methods and have significant error bars.  In this work, we introduce the trimmed sampling algorithm in order to solve this problem.  Using the framework of Bayesian inference, we sample prior probability distributions determined by uncertainty estimates of the various matrix elements and likelihood functions composed of physics-informed constraints.  The result is a probability distribution for the eigenvectors and observables which automatically comes with a reliable estimate of the error and performs far better than standard regularization methods.  The method should have immediate use for a wide range of applications involving classical and quantum computing calculations of large quantum systems.
\end{abstract}

\maketitle

\section{Introduction}
One common approach for finding extremal eigenvalues and eigenvectors of large quantum systems is to project onto basis states that have good overlap with the eigenvector of interest.  Since these states are often not orthogonal to each other, this process results in a generalized eigenvalue problem of the form $H\ket{\psi}=EN\ket{\psi}$, where $H$ is the projected Hamiltonian matrix, $N$ is the norm matrix for the non-orthogonal basis, $E$ is the energy, and $\ket{\psi}$ is the column vector for the projected eigenvector.  If $O$ is the projected matrix for some other observable using the same basis, then we can compute expectation values of that observable using $\braket{O} = \braket{\psi | O | \psi }/\braket{\psi | N | \psi}$.

The generator coordinate method is a well-known technique in nuclear physics, where the corresponding generalized eigenvalue problem is called the Hill-Wheeler equation \cite{Hill:1952jb,Griffin:1957zza,Wong:1975,Magierski:1995zz}.  The generalized eigenvalue problem is used in several computational approaches utilizing variational subspace methods and non-orthogonal bases \cite{Varga:1995dm,Blume:2009,Togashi:2018gkq}.  It serves as a cornerstone of methods such as eigenvector continuation \cite{Frame:2017fah,Demol:2019yjt,Konig:2019adq,Ekstrom:2019lss,Furnstahl:2020abp,Sarkar:2020mad} and the more general class of reduced basis methods \cite{quarteroni2015reduced,Bonilla:2022rph,Melendez:2022kid}.  It is also useful for Monte Carlo simulations where trial states are produced using Euclidean time projection starting from several different initial and final states \cite{HadronSpectrum:2012gic,Edwards:2012fx,Elhatisari:2019,Shen:2021kqr,Shen:2022bak}.

By using only a small subspace of states, the generalized eigenvalue problem can make efficient use of computational resources.  However, one major weakness is the sensitivity to error. One needs to find eigenvectors and eigenvalues of the matrix $N^{-1}H$, and the condition number of the norm matrix grows larger as the size of the subspace grows.  Even small errors due to the limits of machine precision can cause problems.  But the problem is even worse for stochastic methods.
Monte Carlo simulations are among the most powerful tools for solving quantum many body systems.  Unfortunately, Monte Carlo calculations produce statistical errors when calculating elements of the Hamiltonian and norm matrices, and the resulting uncertainties for the generalized eigenvalue problem can be very large.  The same challenges arise for quantum computing where all measurements are stochastic in nature.  When using variational subspace methods in quantum computing \cite{Francis:2022zib}, one must consider both statistical errors as well as systematic errors due to gate errors, measurement errors, and decoherence \cite{Peruzzo:2014,Dumitrescu:2018njn,Qian:2021wya,Bee-Lindgren:2022nqb}. 

There are well-established methods for dealing with ill-posed inverse problems. Tikhonov regularization is one popular approach \cite{Tikhonov:1943}, and the simplest form of Tikhonov regularization is ridge regression or nugget regularization. In this approach a small positive multiple of the identity, $\epsilon I$, is added to the norm matrix that needs to be inverted.  However, it is not straightforward to choose an appropriate value for $\epsilon$ \cite{Alheety:2011} nor to estimate the systematic bias introduced by the regularization.

In this work we introduce the trimmed sampling algorithm, which uses physics-based constraints and Bayesian inference \cite{Gelman:1995,Phillips:2020dmw} to reduce errors of the generalized eigenvalue problem.  Instead of simply regulating the norm matrix, we sample probability distributions for the Hamiltonian and norm matrix elements weighted by likelihood functions derived from physics-informed constraints about positivity of the norm matrix and convergence of extremal eigenvalues with respect to subspace size.  We determine the posterior distribution for the Hamiltonian and norm matrix elements and sample eigenvectors and observables from that distribution.  To demonstrate the method, we apply the trimmed sampling algorithm to two challenging benchmark calculations.   We analyze the performance and discuss new applications that may be possible based upon this work and its extensions.

\section{Methods}
Let us label the basis states used to define the generalized eigenvalue problem as $\ket{v_1}, \ket{v_2}, \cdots \ket{v_n}, \cdots$.  As noted above, there is no assumption that the states are orthogonal or normalized.  Let us define $E_n$ as the ground state energy for the generalized eigenvalue problem if we truncate after the first $n$ basis states,
\begin{equation}
    H^{(n \times n)} \ket{\psi_n} = E_n N^{(n\times n)} \ket{\psi_n}.\label{eq:energies}
\end{equation}
If the matrix elements of $H$ and $N$ could be determined exactly, then the variational principle tells us that the sequence $E_n$ must be monotonically decreasing and bounded below by the true ground state energy $E_{\rm exact}$.  We will assume that the basis states $\ket{v_1}, \ket{v_2}, \cdots \ket{v_n}, \cdots$ have been ordered so that the energies $E_n$ converge to $E_{\rm exact}$ as a smoothly-varying function of $n$ for sufficiently large $n$. 

The problem is that we do not have exact calculations of the $H$ and $N$ matrices.  Instead we start with some estimates for the Hamiltonian and norm matrices, which we call $\tilde{H}$ and $\tilde{N}$  respectively.  We also are given one-standard-deviation error estimates for each element of the Hamiltonian and norm matrices, which denote as $\Delta\tilde{H}$ and $\Delta\tilde{N}$ respectively.  In this work we consider the standard case where the Hamiltonian and norm matrices are manifestly Hermitian.  But we also discuss the generalization to the non-Hermitian case at the end of our analysis.

We will compute a posterior probability distribution $P(H,N|R)$ for the elements of the Hamiltonian matrix $H$ and norm matrix $N$.  Here $R$ indicates a set of physics-informed conditions we impose on the $H$ and $N$ matrices, and the corresponding likelihood function is written as $P(R|H,N)$.  We also include a prior probability distribution, which we write as $P(H,N)$.  From Bayes' theorem, the posterior distribution is given by
\begin{equation}
    P(H,N|R) = \frac{P(R|H,N)P(H,N)}{\int \prod_{ij} [dH_{ij} dN_{ij}] P(R|H,N)P(H,N)}.
    \label{eq:Bayes}
\end{equation}

For our prior distribution, we take a product of uncorrelated Gaussian functions, 
\begin{equation}
    P(H,N) = \prod_{ij} \frac{e^{-\tfrac{(H_{ij}-\tilde{H}_{ij})^2}{2(\Delta \tilde{H}_{ij})^2}}e^{-\tfrac{(N_{ij}-\tilde{N}_{ij})^2}{2(\Delta \tilde{N}_{ij})^2}}}{2\pi \Delta \tilde{H}_{ij} \Delta \tilde{N}_{ij}},
\end{equation}
though a more detailed model of the prior distribution with asymmetric errors and correlations among matrix elements can also be implemented.

Our likelihood function $P(R|H,N)$ is a product of two factors, 
\begin{equation}
    P(R|H,N) = \alpha f_{\rm pos}(N) f_{C}(H,N).
\end{equation}
The first factor, $\alpha$, is a normalization constant that cancels in Eq.~(\ref{eq:Bayes}). The second factor, $f_{\rm pos}(N)$, enforces the constraint that the norm matrix must be positive definite.  It equals $1$ if $N$ is positive definite and equals $0$ otherwise.  The final factor $f_{C}(H,N)$ is a function of the submatrix energies $E_n$ given in Eq.~(\ref{eq:energies}).  Let us define the convergence ratio $C_n$ for $n>2$ as 
\begin{equation}
    C_n = \frac{E_n - E_{n-1}}{E_{n-1} - E_{n-2}}.
\end{equation}
We have taken ratios of energy differences for consecutive energies $E_n$.  This can be generalized to ratios of energy differences between non-consecutive energies in cases where the convergence pattern has some periodicity.  Let $C_{\rm max}$ be the maximum of $C_n$ over all $n$.  We define $C$ to be a conservative upper bound estimate for $C_{\rm max}$.  We then take the second likelihood factor, $f_{C}(H,N)$, to have the form 
\begin{equation}
f_{C}(H,N)=e^{-\tfrac{C_{\rm max}}{C}}. \label{eq:likelihood_factor}
\end{equation}
The purpose of this likelihood function is to penalize the likelihood of Hamiltonian and norm matrices whose convergence rate for the ground state energies is much slower than expected and $C_{\rm max}$ is significantly larger than $C$.  Neither the exact value for $C$ nor the exact functional form for $f_{C}(H,N)$ are essential features that need to be finely tuned.  Similar results can be obtained using a wide range of different choices, and for some applications a different definition for $f_{C}(H,N)$ may prove to be more effective.

In order to sample the posterior distribution in $P(H,N|R)$ in Eq.~(\ref{eq:Bayes}), we first produce random samples for the Hamiltonian and norm matrices using a heat bath algorithm given by the prior probability distribution $P(H,N)$.  We then reweight the samples according to the likelihood function $P(R|H,N)$.  From this sampling of the posterior distribution, we can compute weighted median values and estimated error bars for the energies or any other observable.  This importance reweighting scheme is commonly used in Markov Chain Monte Carlo algorithms \cite{Hastings:1970aa}.  It is also similar to the sampling/importance resampling method described in  Ref.~\cite{Smith:1992,Jiang:2022off}, except that we are not resampling data.

\section{Bose-Hubbard Model}

\begin{figure*}
    \centerline{\includegraphics[width=23cm]{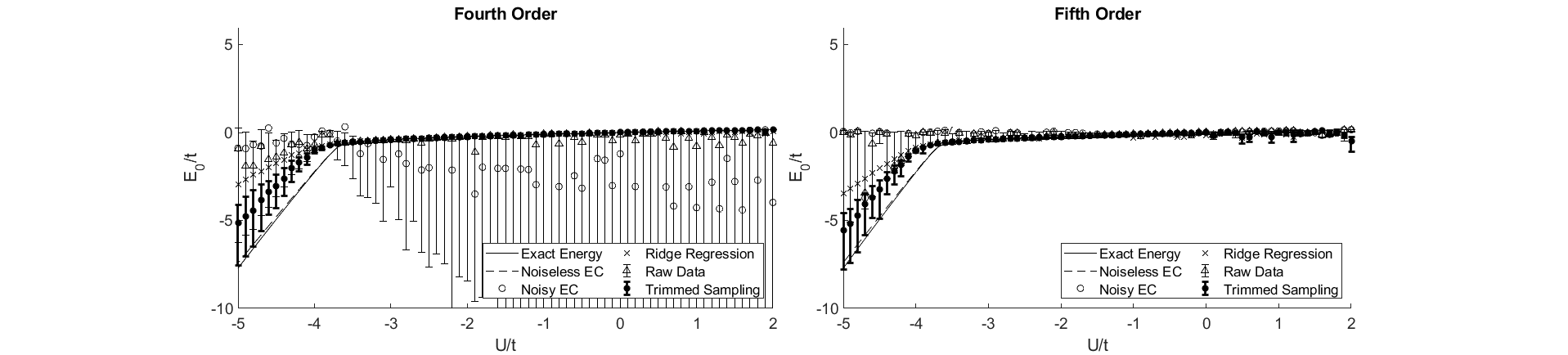}}
    \caption{Ground state energy of the Bose-Hubbard model as a function of coupling strength $U/t$. The ``exact'' ground state energies are plotted as solid lines.  The ``noiseless EC'' data are presented with dashed lines.  The ``noisy EC'' results corresponding with matrix elements $\tilde{H}$ and $\tilde{N}$ are plotted with open circles.  The results using ``ridge regression'' are shown with times symbols.  The ``raw data'' obtained by sampling the prior probability distribution are displayed with open triangles and error bars.  The ``trimmed sampling'' results are plotted as filled circles with error bars.}
    \label{fig:BoseHubbard}
\end{figure*}

For the first benchmark test of the trimmed sampling algorithm, we consider the Bose-Hubbard model in three-dimensions. This system describes a system of identical bosons on a three-dimensional lattice, with a Hamiltonian that contains a hopping term proportional to $t$, a contact interaction proportional to $U$, and a chemical potential proportional to $\mu$. We will consider the system with four bosons on a $4\times4\times4$ lattice with $\mu=-6t$.  Further details of the model can be found in the Supplemental Material.  Following the analysis in Ref.~\cite{Frame:2017fah}, we use eigenvector continuation (EC) to determine the ground state energy for a range of couplings $U/t$.  For our basis states we use the ground state eigenvectors for five training values, $U/t=-2.5,-1.9,-1.8,-1.7,-1.6$.

In order to introduce noise into the EC calculations, we round each entry of the $H$ and $N$ matrices at the sixth decimal place and use these rounded values for our estimates $\tilde{H}$ and $\tilde{N}$.  Since the rounding error is performed at the sixth digit, we use the error estimates $\Delta \tilde{H}_{ij} = \Delta \tilde{N}_{ij} = \tfrac{1}{\sqrt{12}} \times 10^{-6}$. While a uniform error distribution more accurately captures the nature of this error, for this example we assume Gaussian noise to demonstrate that it is not necessary to know the exact form of the errors. For our $n^{\rm th}$ order EC calculation, we use the first $n$ basis states and apply trimmed sampling. 

The results for the ground state energies $E_0/t$ versus coupling $U/t$ are presented in Fig.~\ref{fig:BoseHubbard} for orders $n = 4,5$.  The ``exact'' ground state energies are shown with solid lines.  The ``noiseless EC'' results are plotted with dashed lines.  The ``noisy EC'' results corresponding with matrix elements $\tilde{H}$ and $\tilde{N}$ are displayed with open circles.  The results obtained using ``ridge regression'' are plotted with times symbols. We have optimized the parameter $\epsilon$ used in ridge regression by hand to produce results as close as possible to the noiseless EC results.  All other calculations using ridge regression will therefore not be better than the idealized ridge regression results we present.  The ``raw data'' obtained by sampling the prior probability distribution $P(H,N)$ associated with $\tilde{H}$ and $\tilde{N}$ and uncertainties $\Delta \tilde{H}$ and $\Delta \tilde{N}$ are displayed with open triangles and error bars.

The ``trimmed sampling'' results are obtained by sampling the posterior probability distribution $P(H,N|R)$ and plotted as filled circles with error bars.  For all of our plots showing error bars, the plot symbol is located at the weighted median value while the lower and upper limits correspond to the $16^{\rm th}$ and $84^{\rm th}$ percentiles respectively.   We find that this representation of the error bars is useful since the distributions have much heavier tails than Gaussian distributions.  For all of the trimmed sampling results present here, we have produced $500$ samples with nonzero posterior probability, and the small matrix calculations can be performed easily on a single processor.

We use the value $C = 2.5$ for $f_{C}(H,N)$ in Eq.~(\ref{eq:likelihood_factor}).  The trimmed sampling algorithm is clearly doing a good job of controlling errors due to noise.  The trimmed sampling algorithm is performing significantly better than the standard regularization provided by ridge regression.  There is some systematic underestimation of the error near the avoided level crossing at $U/t = -3.8$.  Overall, however, the trimmed sampling error bar gives a reasonable estimate of the actual deviation from the actual noiseless EC results.  As discussed in the Supplemental Material, the trimmed sampling error bars correspond to the distribution of values obtained for the observable of interest while sampling the posterior probability distribution.  While this is not an unbiased estimate, it does serve as an approximate estimate of the actual error in the sense that the exact result is a point in the posterior distribution with non-negligible weight.

\begin{figure*}
    \centerline{\includegraphics[width=23cm]{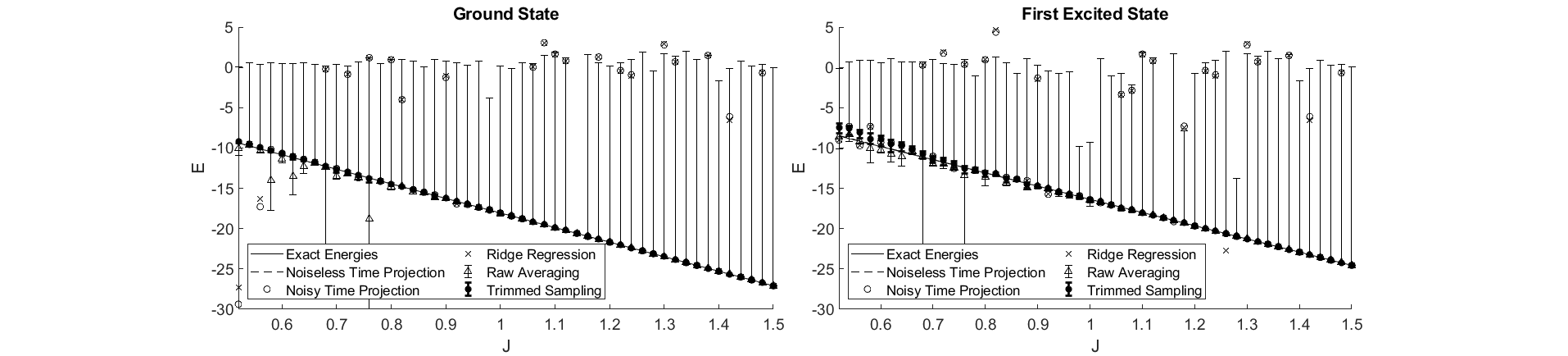}}
    \caption{Ground state and first excited state energies of the one-dimensional Heisenberg chain as a function of coupling strength $J$. The ``exact'' energies are plotted as solid lines.  The ``noiseless time projection'' data are dashed lines.  The ``noisy time projection'' results corresponding with matrix elements $\tilde{H}$ and $\tilde{N}$ are plotted with open circles.  The data obtained using ``ridge regression'' are shown with times symbols.  The ``raw data'' obtained by sampling the prior probability distribution are drawn with open triangles and error bars.  The ``trimmed sampling'' results are plotted as filled circles with error bars. }
    \label{fig:TPHistogram}
\end{figure*}

\section{Heisenberg Model}
For the second benchmark test, we consider a one-dimensional quantum Heisenberg chain. The Hamiltonian for this system is
\begin{equation}
    H_J = -J\sum_{j=1}^{N}[\sigma^x_{j}\sigma^x_{j+1}+\sigma^y_{j}\sigma^y_{j+1}+\sigma^z_{j}\sigma^z_{j+1}].
\end{equation}
Here $\sigma^x, \sigma^y, \sigma^z$ are the Pauli matrices, $N$ is the number of sites, and $J$ is the coupling. We consider the case with $N=10$ sites and calculate the lowest four energy eigenvalues of the subspace with $\sum_j \sigma^z_{j}=0$.   For more details of the model, see Ref.~\cite{Choi:2021}.

For the generalized eigenvalue problem, we construct our basis states using Euclidean time projection, starting from the initial state $\ket{v_0}=\ket{0101010101}$.  We are using the standard qubit notation where $\ket{0}$ is the $+1$ eigenstate of $\sigma_z$ and $\ket{1}$ is the $-1$ eigenstate of $\sigma_z$.  We operate on $\ket{v_0}$ with the Euclidean time projection operator, $e^{-H t}$.  This is equivalent to how projection Monte Carlo simulations are performed \cite{Lee:2008fa,Lahde:2019npb}.  We consider values of $J$ ranging from $0.5$ to $1.5$.  For each value of $J$, we take the Euclidean time values $t_n=0.1 n$ and define each basis vector as $\ket{v_n}=e^{-H t_n}\ket{v_0}$ for $n = 0,1,2,3,4,5$. After projecting onto these six vectors, we calculate the corresponding Hamiltonian and norm matrices and solve the generalized eigenvalue problem.   

In order to introduce noise into the calculation, we apply random Gaussian noise with standard deviation $\sigma=0.01$ to each element of the Hamiltonian and norm matrices. These resulting matrices with noise define our estimates $\tilde{H}$ and $\tilde{N}$, and we take the uncertainty estimates to be $\Delta \tilde{H}_{ij} = \Delta \tilde{N}_{ij} = 0.01$.  For each value of $J$, the observables we compute are the lowest four energy eigenvalues.  Since $J$ is just an overall scale for the Hamiltonian, the exact energies will just scale linearly with $J$.  However, the Euclidean time projection calculations will be different for each $J$ due to the fixed projection times $t_n$ used.  The random noise will also be different for different values of $J$.

The results are presented in Fig.~\ref{fig:TPHistogram}.  The ``exact'' energies calculated using exact diagonalization are plotted with solid lines.  The ``noiseless time projection'' results are shown with dashed lines.  The ``noisy time projection'' results corresponding with matrix elements $\tilde{H}$ and $\tilde{N}$ are plotted with open circles.  The results obtained using ``ridge regression'' are displayed with times symbols.  We have again optimized the parameter $\epsilon$ used in ridge regression to produce the best possible performance, though the overall improvement is not significant. The ``raw data'' obtained by sampling the prior probability distribution $P(H,N)$ associated with mean values $\tilde{H}$ and $\tilde{N}$ and uncertainties $\Delta \tilde{H}$ and $\Delta \tilde{N}$ are presented with open triangles and error bars.

The ``trimmed sampling'' results are obtained by sampling the posterior probability distribution $P(H,N|R)$ and plotted as filled circles with error bars. These results use the value $C = 2.5$ for $f_{C}(H,N)$ in Eq.~(\ref{eq:likelihood_factor}). The trimmed sampling algorithm is again doing a good job of controlling errors due to noise, and the trimmed sampling error bars give a reasonable estimate of the actual deviation from the ``noiseless time projection'' results.  In contrast, ridge regression is not giving consistently reliable results for this benchmark test.
\begin{figure*}
    \centerline{\includegraphics[width=23cm]{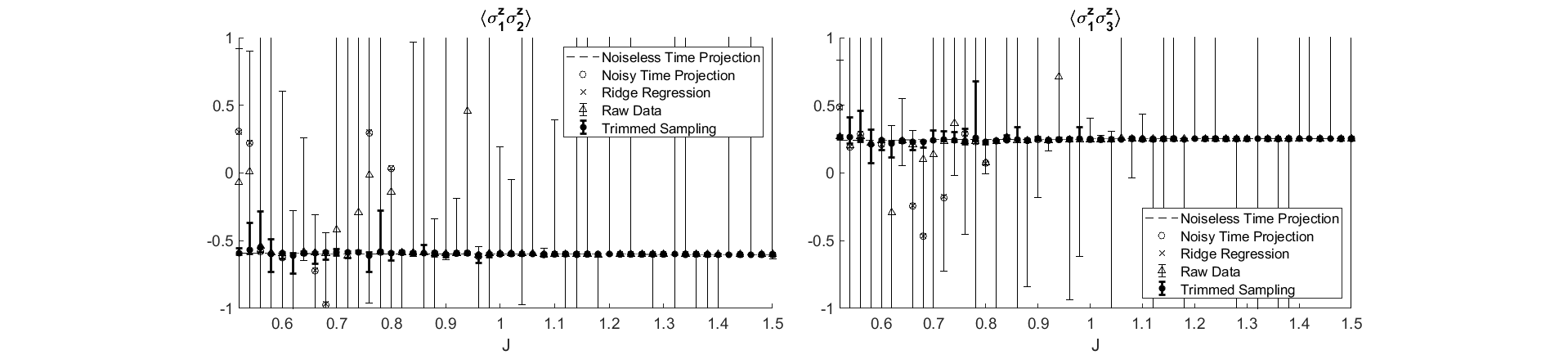}}
    \caption{Spin pair expectation values for the ground state of the Heisenberg model as a function of $J$.  $\braket{\sigma_1^z \sigma_2^z}$ is shown in the left panel, and $\braket{\sigma_1^z \sigma_3^z}$ is presented in the right panel.  The plot symbols are the same as in Fig.~\ref{fig:TPHistogram}.
    \label{fig:Obs_ground}}
\end{figure*}

In addition to calculating energies for the Heisenberg model, we can also compute spin observables.  In Fig.~\ref{fig:Obs_ground} we show results for the ground-state expectation value of the product of nearest-neighbor spins $\braket{\sigma_1^z \sigma_2^z}$ in the left panel and the product of next-to-nearest-neighbor spins $\braket{\sigma_1^z \sigma_3^z}$ in the right panel.  The trimmed sampling algorithm is again performing significantly better than ridge regression. The trimmed sampling error bars also provide a reasonable estimate of the actual deviation from the ``noiseless time projection'' results.

\section{Discussion and Outlook}

The generalized eigenvalue problem is a useful method for finding the extremal eigenvalues and eigenvectors of large quantum systems.  However, the approach is highly susceptible to noise.  We have presented the trimmed sampling algorithm, which uses the framework of Bayes inference to incorporate information about the prior probability distribution for the Hamiltonian and norm matrix elements together with physics-informed likelihood constraints.  The result is a posterior probability distribution that can easily be sampled.  For the benchmark examples presented here, we find that trimmed sampling performs significantly better than standard regularization methods such as ridge regression.  We have demonstrated significant error reductions for energy calculations as well as other observables.  In the Supplemental Material, we present several other benchmark calculations that further demonstrate the performance of the trimmed sampling algorithm.

The trimmed sampling algorithm can be used for any generalized eigenvalue problem obtained using classical computing or quantum computing.  This encompasses a very wide class of problems ranging from quantum many body systems to partial differential equations to quantum field theories.  All that is needed are some good estimates for the Hamiltonian and norm matrix elements and their corresponding uncertainties.  In order to gain the full advantage of the trimmed sampling algorithm, it is important that matrix calculations are performed using machine precision that is finer than the uncertainties of the Hamiltonian and norm matrix elements.  Studies of the trimmed sampling algorithm for the non-Hermitian Hamiltonian and norm matrices are currently under investigation.  In that case one cannot simply impose positivity of the norm matrix in the likelihood function. However, one can instead impose more stringent conditions on the convergence of the energies $E_n$ for the submatrix calculations.

\paragraph*{Acknowledgements}
We are grateful for discussions with Joey Bonitati, Serdar Elhatisari, Gabriel Given, Zhengrong Qian, Avik Sarkar, Jacob Watkins, and Xilin Zhang.   We acknowledge financial support from the U.S. Department of Energy (DE-SC0021152 and DE-SC0013365) and the NUCLEI SciDAC-4 collaboration.
\bibliography{References}

\clearpage
\onecolumngrid

\renewcommand{\thefigure}{S\arabic{figure}}

\setcounter{figure}{0}

\section{Supplemental Materials}

\subsection*{Bose-Hubbard model}
In this work, we consider
a Bose-Hubbard model which consists of an interacting system of identical bosons on a three-dimensional spatial lattice.  The Hamiltonian features a term with hopping coefficient $t$ that controls the nearest-neighbor
hopping of each boson, an interaction coefficient $U$ responsible for pairwise interactions between bosons on the
same lattice site, and
a chemical potential $\mu$.  The Hamiltonian is defined on a three-dimensional spatial cubic lattice and has the form  
\begin{equation}
H = -t \sum_{\langle \bf{n'},\bf{n} \rangle} a^{\dagger}({\bf n'})a({\bf
n}) + 
\frac{U}{2}\sum_{\bf n}\rho(\bf{n})[\rho(\bf{n})-1]-\mu \sum_{\bf n}\rho(\bf{n}),
\end{equation}
where $a({\bf n})$ and $a^{\dagger}({\bf n})$ are the annihilation and creation
operators for bosons at lattice site ${\bf n}$.  The first summation is over
nearest-neighbor pairs
$\langle {\bf n'},{\bf n}\rangle$, and $\rho({\bf n})$ is the density operator
$a^{\dagger}({\bf n})a(\bf{n})$.  For our example, we consider a system
of four bosons with $\mu =
-6t$ on a $4\times 4\times 4$ lattice.  This choice sets the ground state energy for the non-interacting
case $U=0$ equal to zero. 

\subsection{Additional benchmark calculations}

We present some additional benchmark calculations of the trimmed sampling algorithm using the one-dimensional quantum Heisenberg chain presented in the main text.  As described there, we construct basis states using Euclidean time projection, starting from the initial state $\ket{v_0}=\ket{0101010101}$.  We operate on $\ket{v_0}$ with the Euclidean time projection operator, $e^{-H t}$.  For each value of $J$, we take the Euclidean time values $t_n=0.1 n$ and define each basis vector as $\ket{v_n}=e^{-H t_n}\ket{v_0}$ for $n = 0,1,2,3,4,5$. After projecting onto these six vectors, we calculate the corresponding Hamiltonian and norm matrices and solve the generalized eigenvalue problem.  In order to introduce noise into the calculation, we apply random Gaussian noise with standard deviation $\sigma=0.01$ to each element of the Hamiltonian and norm matrices.

\begin{figure*}[h]
    \centerline{\includegraphics[scale=0.45]{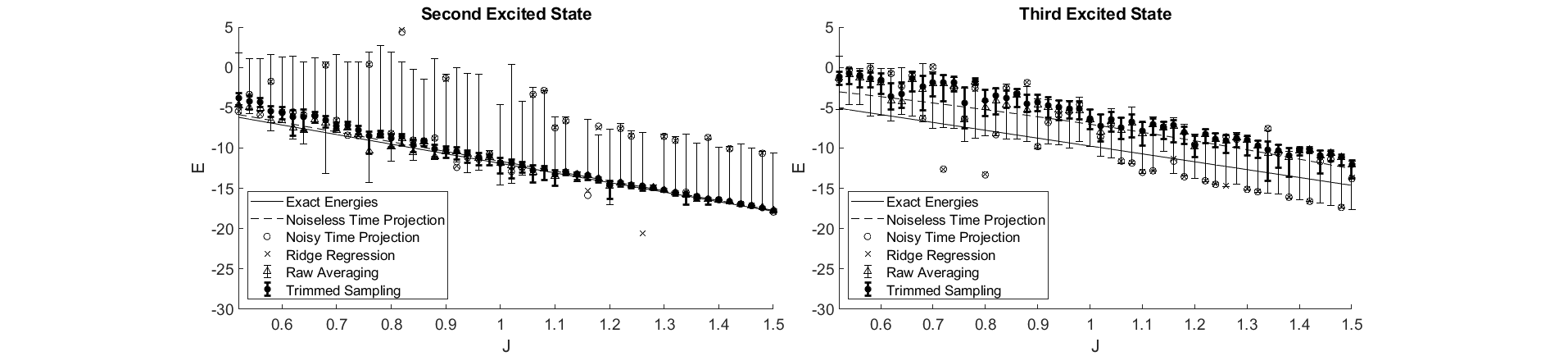}}
    \caption{Second and third excited state energies of the one-dimensional Heisenberg chain as a function of coupling strength $J$. The ``exact'' energies are plotted as solid lines.  The ``noiseless time projection'' data are dashed lines.  The ``noisy time projection'' results corresponding with matrix elements $\tilde{H}$ and $\tilde{N}$ are shown with open circles.  The data obtained using ``ridge regression'' are plotted with times symbols.  The ``raw data'' obtained by sampling the prior probability distribution are drawn with open triangles and error bars.  The ``trimmed sampling'' results are plotted as filled circles with error bars. }
    \label{fig:Heisenberg_energies_2_3}
\end{figure*}

In Fig.~\ref{fig:Heisenberg_energies_2_3}, we show results for the energies of the second excited state and third excited state of the Heisenberg model.  The ``exact'' energies are calculated using exact diagaonalization are shown with solid lines.  The ``noiseless time projection'' results are plotted with dashed lines.  The ``noisy time projection'' results corresponding with matrix elements $\tilde{H}$ and $\tilde{N}$ are plotted with open circles.  The results obtained using ``ridge regression'' are displayed with times symbols.  We have tried to optimize the parameter $\epsilon$ used in ridge regression to produce the best possible performance, though the overall improvement is not significant.  The ``raw data'' obtained by sampling the prior probability distribution $P(H,N)$ associated with mean values $\tilde{H}$ and $\tilde{N}$ and uncertainties $\Delta \tilde{H}$ and $\Delta \tilde{N}$ are presented with open triangles and error bars.  The ``trimmed sampling'' results are plotted as filled circles with error bars.  These results use the value $C = 2.5$ for $f_{C}(H,N)$.  We see that the trimmed sampling algorithm is performing significantly better than ridge regression. The trimmed sampling error bars give a reasonable estimate of the actual deviation from the ``noiseless time projection'' results.
     
\begin{figure*}
    \centerline{\includegraphics[scale=0.45]{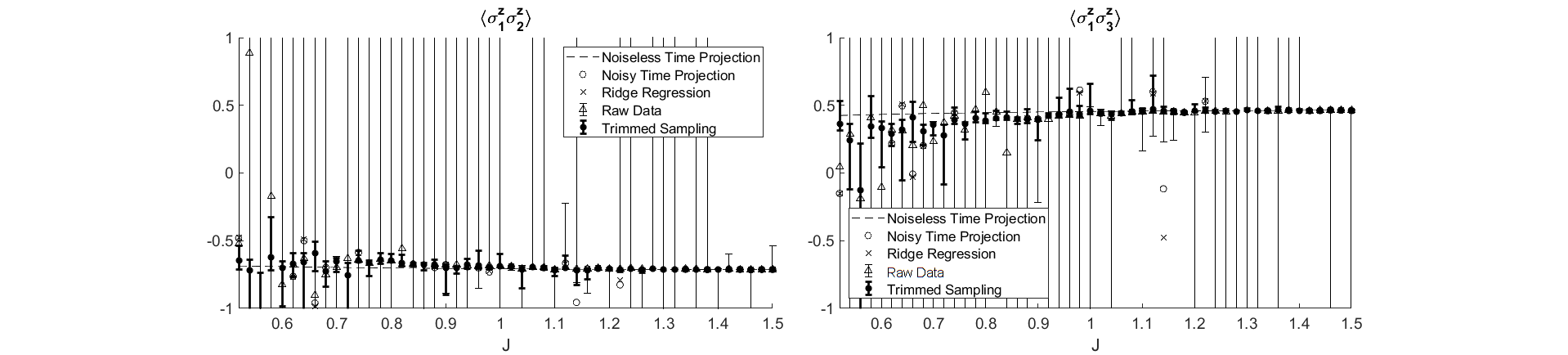}}
    \caption{Spin pair expectation values for the first excited state of the Heisenberg model as a function of $J$.  $\braket{\sigma_1^z \sigma_2^z}$ is presented in the left panel and $\braket{\sigma_1^z \sigma_3^z}$ is shown in the right panel.  The plot symbols are the same as in Fig.~\ref{fig:Heisenberg_energies_2_3}.
    \label{fig:Heisenberg_spins_1}}
\end{figure*}

In Fig.~\ref{fig:Heisenberg_spins_1} we show results for spin pair expectation values for the first excited state of the Heisenberg model.  The product of nearest-neighbor spins $\braket{\sigma_1^z \sigma_2^z}$ is shown in the left panel, and the product of next-to-nearest-neighbor spins $\braket{\sigma_1^z \sigma_3^z}$ is plotted in the right panel.  These results use the value $C = 2.5$ for $f_{C}(H,N)$.  The trimmed sampling algorithm is once again performing better than ridge regression.  The trimmed sampling error bars give a reasonable estimate of the actual deviation from the ``noiseless time projection'' results.

\subsection{Trimmed sampling error estimates}
\begin{figure}
    \centering
    \includegraphics[scale=0.5]{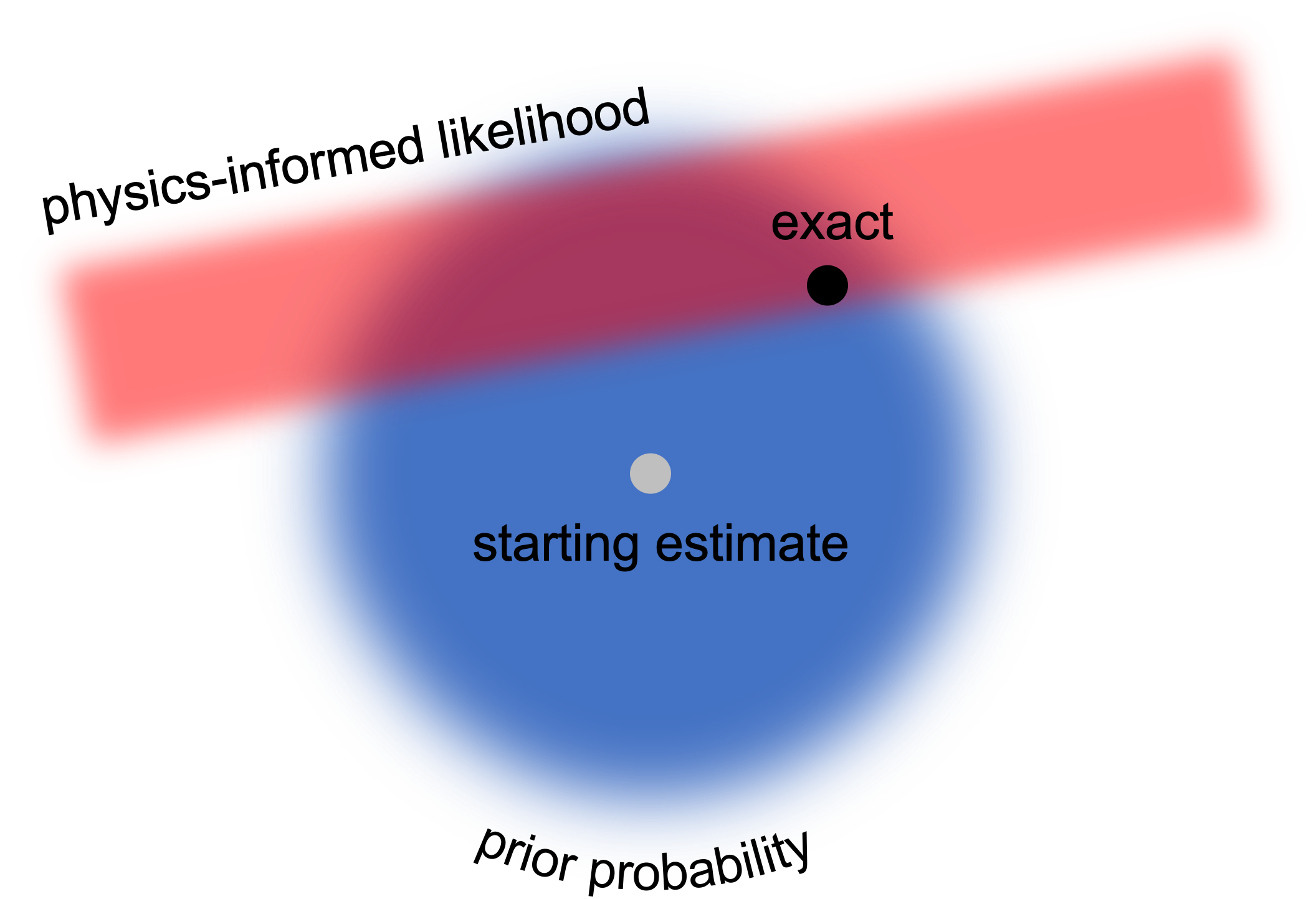}
    \caption{In this schematic diagram, we show the prior probability distribution centered around the starting estimate as well as the physics-informed likelihood. The posterior probability is proportional to the product of the prior probability and the likelihood.  The exact solution is located at a point where both the prior probability and the likelihood are not small.}
    \label{fig:my_label}
\end{figure}

The trimmed sampling error bars we report in this work correspond to the distribution of values obtained for the observable of interest while sampling the posterior probability distribution.  As illustrated in Fig.~\ref{fig:my_label}, the posterior probability is proportional to the product of the prior probability and the likelihood.  The posterior probability does not give an unbiased estimate for the exact value of the observable.  Such an unbiased estimate would be difficult to obtain without much more detailed information about the properties of the system at hand.  However, it can be said that the exact solution to the generalized eigenvalue problem is located at a point where both the prior probability and the likelihood are not small.  So the posterior probability error bar does serve as an approximate estimate of the actual error in the sense that the exact result is a point in the posterior distribution with non-negligible weight.
\subsection{Sensitivity studies}
\begin{figure*}
    \centerline{\includegraphics[scale=0.45]{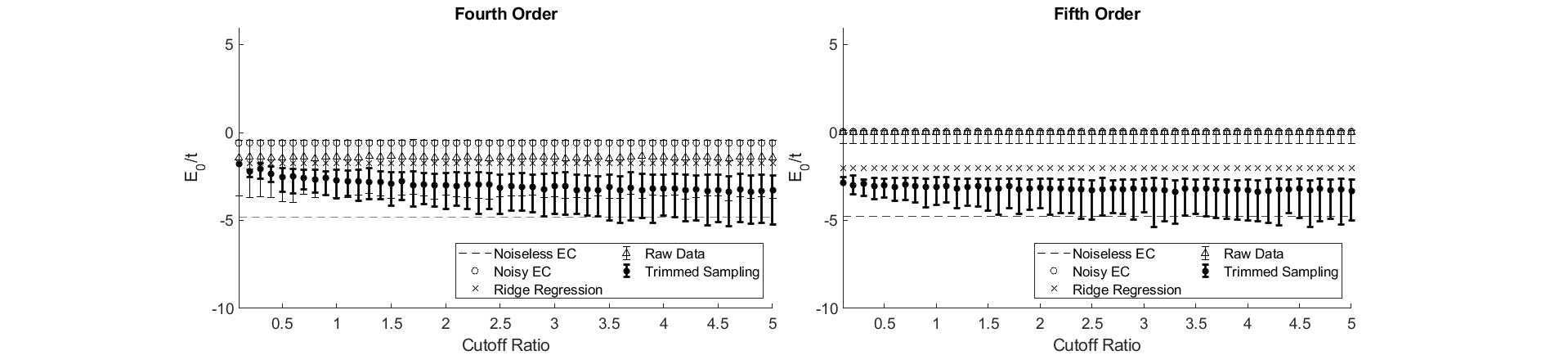}}
    \caption{Comparison of various choices of the cutoff ratio, $C$, for the Bose-Hubbard model. Results for fourth order calculations are presented in the left panel, and results for fifth order calculations are presented in the right panel.
    \label{fig:Cutoff_Variation}}
\end{figure*}

For the benchmark examples presented in the main text, we chose likelihood functions that are simple but which also might seem arbitrary.  Here we present several sensitivity studies which show that the trimmed sampling algorithm is largely insensitive to the details of the likelihood function.  In Fig.~\ref{fig:Cutoff_Variation} we show the results obtained while varying the convergence ratio parameter $C_n$ over the range from 0.02 to 5 for the Bose-Hubbard model for the coupling strength $U/t=-4$. We choose this coupling strength as it lies beyond the avoided level crossing and we observe large differences between the methods presented. We see that the value of the cutoff ratio has limited impact on the results of the trimmed sampling, provided it is not overly restrictive. In cases where $C_n$ is very small, we are systematically biasing the posterior probability so much that the reported error bars are significantly smaller than the deviation from the exact result.

\begin{figure*}
\centerline{\includegraphics[scale=0.45]{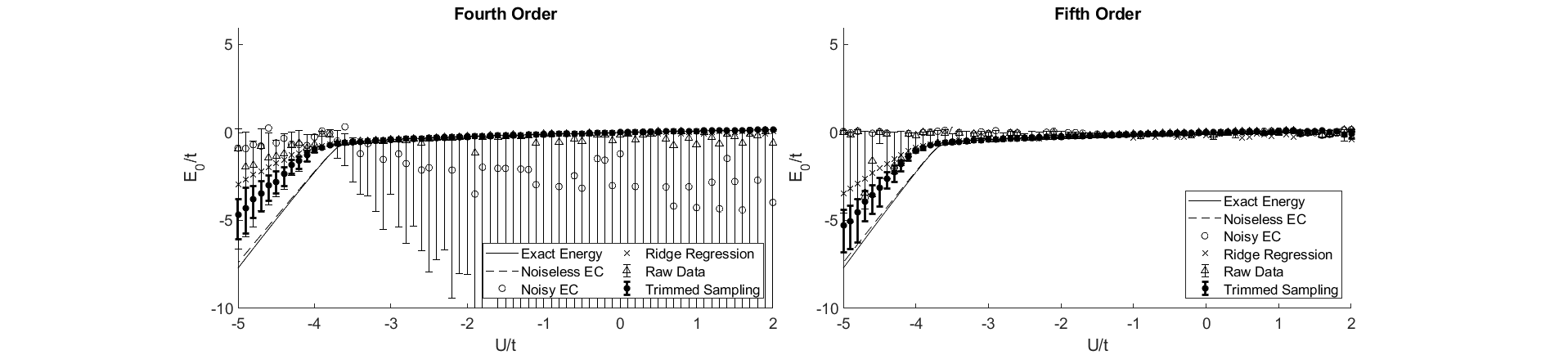}}
    \caption{Calculation of energies for the Bose-Hubbard model using a Gaussian form for the likelihood function. Results for fourth order calculations are presented in the left panel, and results for fifth order calculations are presented in the right panel.    
    \label{fig:GaussianLikelihood}}
\end{figure*}

\begin{figure*}
    \centerline{\includegraphics[scale=0.45]{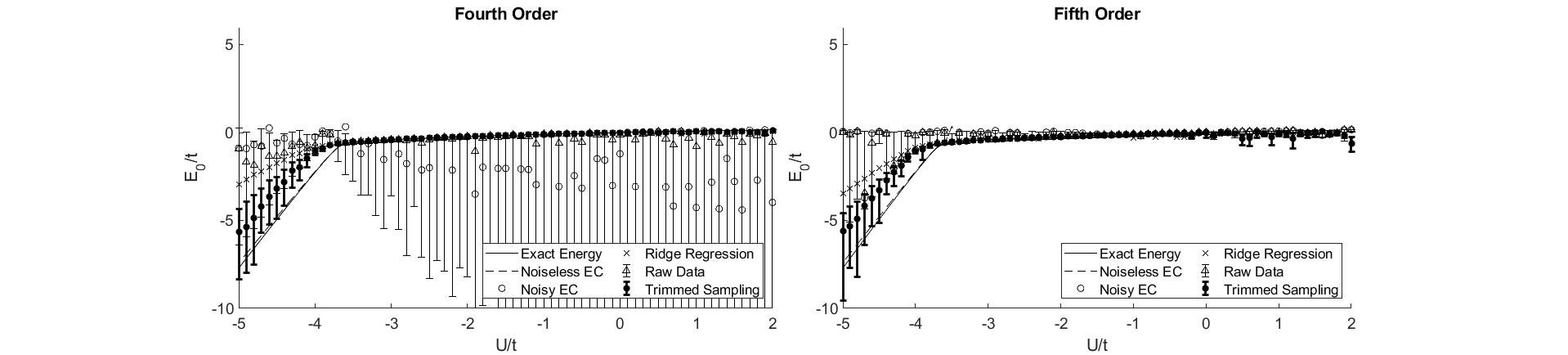}}
    \caption{Calculation of energies for the Bose-Hubbard model using a linear-times-exponential form for the likelihood function. Results for fourth order calculations are presented in the left panel, and results for fifth order calculations are presented in the right panel.
    \label{fig:PoissonLikelihood}}
\end{figure*}

\begin{figure*}
    \centerline{\includegraphics[scale=0.45]{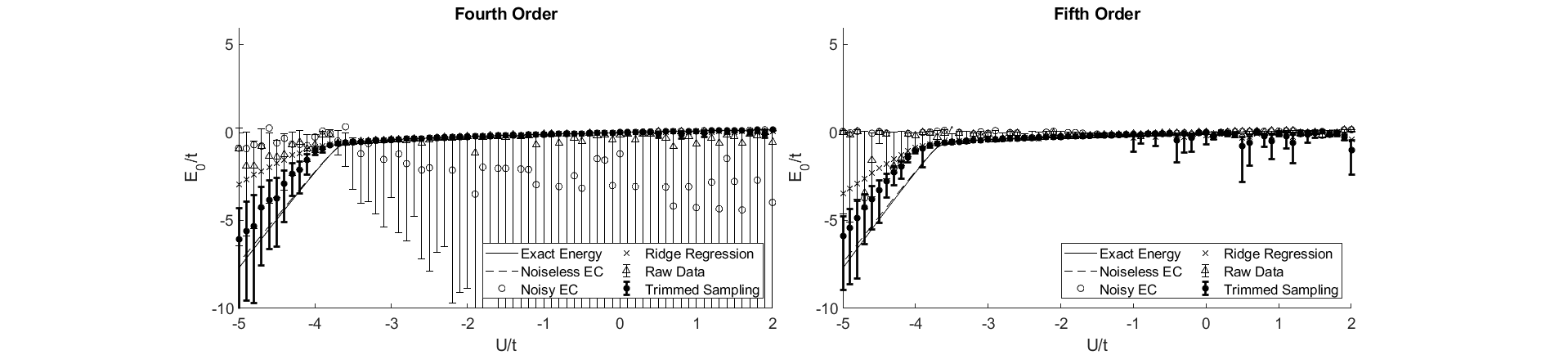}}
    \caption{Calculation of energies for the Bose-Hubbard model using a rational function form for the likelihood function. Results for fourth order calculations are presented in the left panel, and results for fifth order calculations are presented in the right panel. \label{fig:ReciprocalLikelihood}}
\end{figure*}

\begin{figure*}
    \centerline{\includegraphics[scale=0.45]{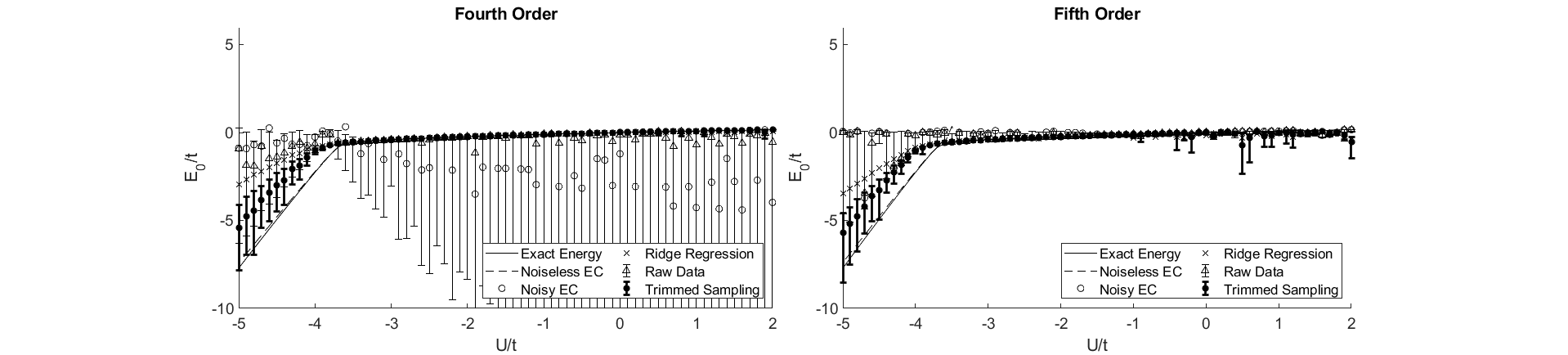}}
    \caption{Calculation of energies for the Bose-Hubbard model using a rectified sinc function form for the likelihood function. Results for fourth order calculations are presented in the left panel, and results for fifth order calculations are presented in the right panel. \label{fig:SincLikelihood}}
\end{figure*}

In Figs.~\ref{fig:GaussianLikelihood}, \ref{fig:PoissonLikelihood}, \ref{fig:ReciprocalLikelihood}, and \ref{fig:SincLikelihood}, we present results for the Bose-Hubbard model using different functional forms for the likelihood function factor presented in main text. We see in all these examples that the functional form does not substantially affect the performance of trimmed sampling.  In Fig.~\ref{fig:GaussianLikelihood}, we used a Gaussian likelihood factor of the form 
\begin{equation}
f_{C}(H,N)=e^{-\tfrac{C^2_{\rm max}}{C^2}}.
\end{equation}
All other parameters of the trimmed sampling are left identical to those presented in the main body of this letter. The 14th and 86th percentile marks shift somewhat relative to the exponential form, but the results are qualitatively the same. This Gaussian form goes 0 much faster than the exponential form for large $C_{\max}$. This results in a narrower distribution of accepted matrix pairs, and the reported error bars are somewhat smaller than the deviation from the exact results.

In Fig.~\ref{fig:PoissonLikelihood}, we use a linear-times-exponential likelihood factor of the form
\begin{equation}
f_{C}(H,N)=C_{\rm max} e^{-\tfrac{C_{\rm max}}{C}}.
\end{equation}
All other parameters of the trimmed sampling are left identical to those presented in the main body of this letter. Again in this case, we see that the 14th and 86th percentile marks have shifted, in this case widening. This functional form has a peak at $C$ rather than 0, as the other functions presented here do. Overall, its behavior is somewhat worse than the previous examples, but it still shows a substantial improvement over the noisy case and ridge regression. A function of this form makes little sense in this context, but is presented only to demonstrate that even a poorly suited functional form shows substantial improvement. 

In Fig.~\ref{fig:ReciprocalLikelihood}, we use a rational function of the form
\begin{equation}
f_{C}(H,N)=\frac{1}{1+\tfrac{C_{\rm max}}{C}}.
\end{equation}
All other parameters of the trimmed sampling are left identical to those presented in the main body of this letter. This example is comparable to the others presented here, but as large values of ${c_{\rm max}}$ are not exponentially damped as they are in previous examples, the error bars indicate a wider posterior distribution. Despite this, there is still clear improvement over the noisy EC results and ridge regression.

In Fig.~\ref{fig:SincLikelihood}, we use a rectified sinc function likelihood factor of the form
\begin{equation}
f_{C}(H,N)= \max\left(\frac{\sin{\tfrac{C_{\rm max}}{C}}}{C_{\rm max}},0\right).
\end{equation}
Even though the use of a rectified sinc function makes no sense in this context, we still see significant improvement over other the noisy EC results and ridge regression.

For each of these examples using different likelihood functions, we see that the choice for the likelihood functional form does not significantly alter the results.  The trimmed sampling method improves the calculation of the generalized eigenvalue problem substantially even when the functional forms are rather ill-suited. For the examples in the main text, we choose the exponential form as it is both simple and not overly restrictive, but still exponentially damps large values of ${C_{\rm max}}$.

\end{document}